\documentclass{iau}
\usepackage{natbib}
\usepackage{amsmath}
\usepackage{graphicx}
\usepackage{multirow}
\usepackage{orcidlink}

\begin{document}

\lefttitle{Anders, Queiroz, Malhotra, et al.}
\righttitle{Degeneracies between stellar ages and unresolved multiplicity}

\jnlPage{1}{7}
\jnlDoiYr{2026}
\doival{10.1017/xxxxx}

\aopheadtitle{Proceedings IAU Symposium 408}
\editors{G. Buldgen, A. Vidotto, \& A. Miglio, eds.}

\title{Towards breaking the degeneracy between \\stellar ages and unresolved multiplicity}

\author{Friedrich Anders$^{1,2}$\orcidlink{0000-0003-4524-9363}, Anna B. A. Queiroz$^{3,4}$\orcidlink{0000-0001-9209-7599}, Sagar Malhotra$^{2,1}$\orcidlink{0000-0002-5509-0168}, \\ Cristina Chiappini$^5$\orcidlink{0000-0003-1269-7282}, Arman Khalatyan$^5$\orcidlink{0000-0002-8913-0690}}
\affiliation{$^1$Departament de Física Quàntica i Astrofísica (FQA), Institut de Ciències del Cosmos (ICCUB), Universitat de Barcelona (UB), Martí i Franquès 1, 08028 Barcelona, Spain}
\affiliation{$^2$Institut d'Estudis Espacials de Catalunya (IEEC), Edifici RDIT, Campus UPC, 08860 Castelldefels, Spain}
\affiliation{$^3$Instituto de Astrof\'isica de Canarias, E-38200 La Laguna, Tenerife, Spain}
\affiliation{$^4$Universidad de La Laguna (ULL), Departamento de Astrofísica, 38206, La Laguna, Tenerife, Spain}
\affiliation{$^5$Leibniz-Institut f\"ur Astrophysik Potsdam (AIP),
          An der Sternwarte 16, 14482 Potsdam, Germany}

\begin{abstract}
Stellar age is the central desirable parameter we need to disentangle different Galactic formation scenarios.
Nevertheless, no stellar age dating method is completely model-independent. Even in the best-case scenarios,
present-day stellar age estimates are only precise to $\sim7\%$ and accurate to $\sim10-15\%$. A main drawback of state-of-the-art stellar parameter inference codes like {\tt StarHorse} is that they do not take into account unresolved stellar multiplicity. Especially in the case of nearly equal-mass binaries or higher-order systems on the main sequence, properly taking into account unresolved multiplicity is imperative for the next generation of stellar surveys. Simulation-based inference (SBI) is a robust and flexible approach for dealing with complex models where traditional likelihood-based methods become unfeasible. Here we sketch how SBI can be used to obtain precise stellar ages, masses, binary mass ratios, and other stellar parameters from spectroscopic, photometric, and astrometric measurements. 
\end{abstract}


\maketitle

\section{Introduction}

Many stars, both in the field and in star clusters, reside in multiple systems \citep{Offner2023}. {\it Gaia} \citep{GaiaCollaboration2016, GaiaCollaboration2023V} can resolve millions of wide binaries out to ~1 kpc from the Sun \citep{El-Badry2021, GaiaCollaboration2023A}, but
most multiple systems are either completely or partially unresolved \citep{Fabricius2021}. A main drawback
of state-of-the-art stellar parameter inference codes is that they do not take into account unresolved stellar multiplicity. Especially in the case of nearly equal-mass binaries or higher-order systems on the main sequence (which are quite abundant; \citet{Abt1976, Duquennoy1991, Halbwachs2003, Fuhrmann2017}), we may therefore expect significantly biased results. Unaccounted unresolved binaries lead to higher effective temperatures and higher extinction values \citep{Anders2019, Anders2022}. Properly taking into account unresolved multiplicity is imperative for the next generation of stellar surveys. Other pertinent problems that presently affect stellar age estimation when applied to thousands or millions of stars (even in the best-case scenarios) are:
Inhomogeneities and biases in atmospheric parameters inferred from different spectroscopic surveys (e.g. \citealt{El-Badry2018, Thomas2024}), uncertainties in stellar evolution (such as stellar rotation, mass loss, internal transport mechanisms, [$\alpha$/Fe]-enhancement, or variations in helium abundance; e.g. \citealt{Lebreton2014, Lebreton2014a, Reese2016}), or the variability of the interstellar extinction law (the dependence of dust obscuration on wavelength and
Galactic position; e.g. \citealt{Schlafly2016, Zhang2025}).

\section{Bayesian ischrone fitting codes}

Bayesian isochrone fitting emerged as a natural improvement from the traditional practice of assigning a stellar age by simply finding the closest point on a (single-star) isochrone. Isochrones have strongly non-linear and sometimes overlapping morphology, so observational uncertainties can map into highly asymmetric or multimodal age distributions. \citet{Pont2004} and \citet{Jorgensen2005} formulated stellar-age estimation explicitly in Bayesian terms, using observed spectroscopic stellar parameters such as effective temperature, luminosity, and metallicity. 

\begin{figure}
  \includegraphics[width=\textwidth, trim={0 .4cm 0 4cm},clip]{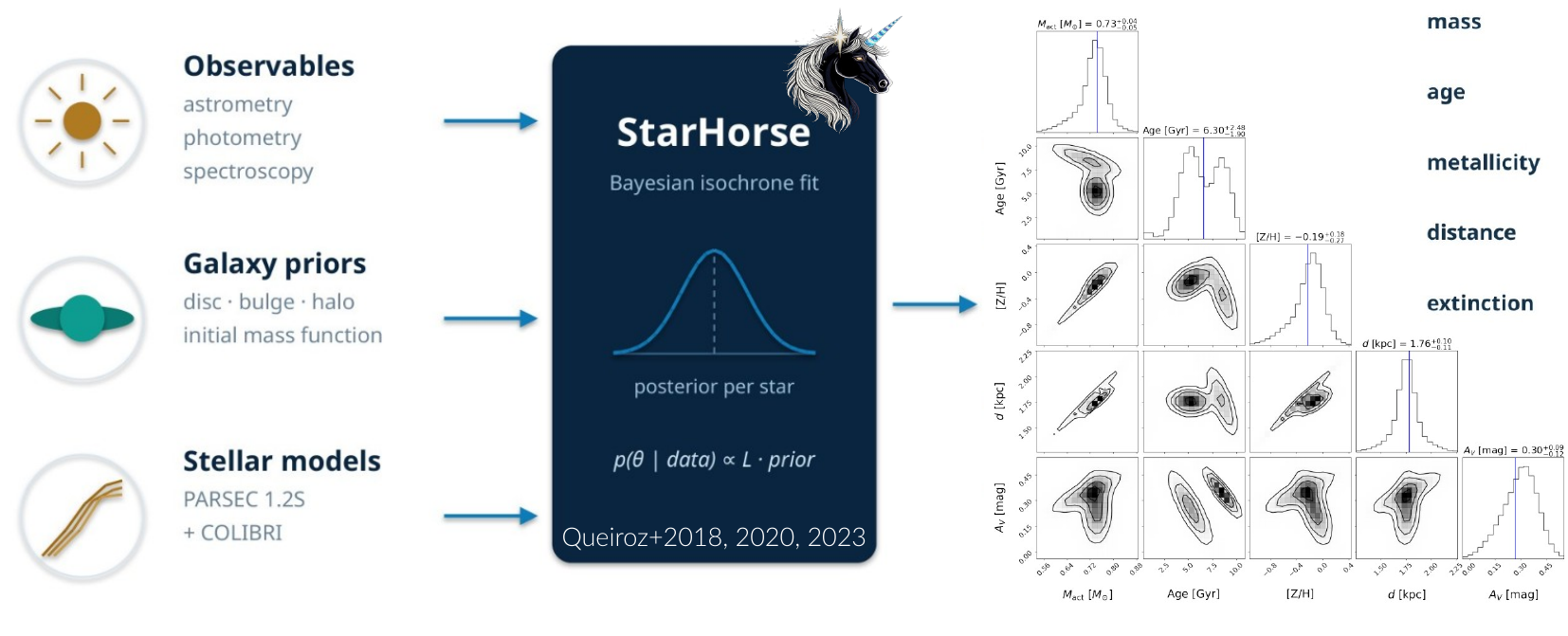}
  \caption{Flowchart of our flexible Bayesian inference code {\tt StarHorse} \citep{Queiroz2018, Queiroz2020, Queiroz2023}. The multimodal posterior shown on the right is an example taken from \citet{Anders2022}, where no spectroscopic input data was used.}
  \label{fig:starhorse_flowchart}
\end{figure}

During the following decade, Bayesian isochrone fitting became a general framework for jointly estimating masses, radii, ages, distances, and related stellar properties, rather than ages alone \citep{daSilva2006, Takeda2007}. Subsequent works incorporated increasingly informative priors, heterogeneous spectroscopic and photometric constraints, and larger stellar-model grids \citep{Burnett2010, Casagrande2011}. 
More recent implementations have extended this philosophy to large-scale spectroscopic surveys \citep[e.g.][]{Binney2014, Santiago2016, Mints2017, Xiang2017, Das2019, Lebreton2020, Kordopatis2023} and to data including asteroseismology \citep[e.g.][]{Rodrigues2017, SilvaAguirre2017, Rendle2019}. In all these works, the conceptual core remains a Bayesian comparison of observations with stellar evolutionary models (e.g. PARSEC, YREC, YY, Dartmouth, or MIST), with increasingly sophisticated likelihoods, priors, (fixed) model physics, and marginalisation over nuisance parameters (see Fig. \ref{fig:starhorse_flowchart}).

The drawback of most of these techniques is that they quickly become very computationally expensive when precise results (i.e. dense parameter grids) are required.
In consequence, machine-learning techniques have emerged as powerful alternatives to these classical methods. For example, \citet{Boin2026} used a surprisingly simple neural-network architecture to predict stellar ages from only three observables (absolute magnitude, reddening-corrected colour, and metallicity), achieving competitive results to classical isochrone fitting (and allowing to quickly compare results from different stellar models). In a similar effort, \citet{Kamulali2025} combined four tree-based machine-learning algorithms to determine ages from only luminosities, effective temperatures, and metallicities, also achieving promising precision and accuracy (compared to asteroseismic benchmark samples). 

Another notable improvement in recent years is the inclusion of full spectral information by several stellar-parameter inference codes (e.g. \citealt{Schonrich2014, Cargile2020, Leung2023, Stone-Martinez2025, Chandra2026, Sun2026}).

\section{Summary of results produced by the {\tt StarHorse} team over the past years}

The {\tt StarHorse} code originated from the need for spectro-photometric (i.e. pre-{\it Gaia}) distances for the Sloan Digital Sky Survey's SEGUE survey \citep{Yanny2009}. 
For \citet{Santiago2016}, the code was translated to {\tt python} and modularised. 
From \citet{Queiroz2018} onward, the {\tt StarHorse} collaboration developed a Bayesian framework for deriving homogeneous stellar distances, extinctions, ages, and fundamental parameters by combining spectroscopic, photometric, and astrometric information. The implementation was designed to infer posterior distributions over stellar evolutionary models while allowing for flexible Galactic priors, incomplete data, and {\it Gaia} parallaxes. We applied it to major spectroscopic stellar surveys (including APOGEE, RAVE, GES, and GALAH), combined these datasets with the latest {\it Gaia} data as well as multi-band photometry, and derived improved distances, extinctions, and stellar parameters for an ever-growing number of objects (500,000 in \citealt{Queiroz2018}, 1.6 million in \citealt{Queiroz2020}, and 11.1 million in \citealt{Queiroz2023} [among them 2.5 million subgiant and main-sequence turn-off stars for which we also publish age estimates]). These value-added catalogues enabled detailed studies of the full extent of the Galactic disc, including its spatial structure, the bar, and the chemical properties of the inner Galaxy. 

In a parallel endeavour, we scaled our methodology to the rapidly expanding {\it Gaia} photo-astrometric data (without using spectroscopy). In \citet{Anders2019, Anders2022} we extended our use of {\tt StarHorse} to objects with only astrometric and photometric observations (combining {\it Gaia}, Pan-STARRS, SkyMapper, 2MASS, and AllWISE), demonstrating that Bayesian isochrone fitting can also provide distances, extinctions, and stellar parameters in this case. The {\it Gaia} EDR3 application produced a catalogue of about 362 million stars brighter than $G=18.5$. 

Finally, \citet{Khalatyan2024} complemented our traditional Bayesian approach by a machine-learning framework ({\tt SHBoost}, based on {\tt XGBoost}; \citealt{Chen2016}) trained on high-quality {\tt StarHorse} results from roughly eight million spectroscopic stars taken from \citet{Queiroz2023}. Using {\it Gaia} DR3 XP spectra, astrometry, and photometry, we extended estimates of extinction and stellar parameters to 217 million stars, demonstrating that the higher precision of the spectroscopic {\tt StarHorse} catalogues could be transferred to the much larger {\it Gaia} XP sample.

\section{A new {\it Gaia} DR3 {\tt StarHorse} run}

The need for a rigorous statistical treatment of stellar-distance inference (see also \citealt{Luri2018}) has recently been highlighted again by \citet{Weiler2025}. He demonstrated that many parallactic distance estimates published over the past decades are affected by a misconception dating back to \citet{Lutz1973}, which has led to the widespread use of an incorrect (and unnecessary) $\propto d^2$ term in the distance prior (e.g. \citealt{Binney2014, Bailer-Jones2015, Santiago2016, Queiroz2018}). 
While this bias affects our latest results for spectroscopic surveys (\citealt{Queiroz2023}) only at a $\sim1\%$ level for ages and distances, the effect becomes much more visible for stars with only photo-astrometric observations and low-signal-to-noise parallax measurements. Especially in the context of Galactic cartography, even small systematic errors in distance estimates can propagate into the identification of spurious structures and lead to biased interpretations of the spatial boundaries of stellar populations.

In our upcoming work (new \texttt{StarHorse} results for \textit{Gaia} DR3 stars), we correct our distance priors following \citet{Weiler2025}. Building upon our previous efforts to derive homogeneous stellar parameters and distances for {\it Gaia} DR3 \citep{Anders2022}, we rerun {\tt StarHorse} for stars with stellar parameter constraints inferred from XP spectra \citep{Khalatyan2024}. 
We will thus provide updated estimates of distances, extinctions, and stellar parameters for approximately 217 million stars.

\section{Revisiting the problem of spectroscopic parameter biases}

In \citet[][Fig. 6]{Queiroz2018}, we tested the impact of systematic atmospheric parameter biases ($T_{\rm eff}, \log g,$, [M/H]) on the output stellar parameters (age, distance, extinction, etc.), based on typical errors reported by spectroscopic stellar surveys at the time. The outcome was (somewhat expectedly) devastating for some of the parameters (see discussion in \citealt{Queiroz2018}, Sect. 4.2) and reiterated the need for accurate calibrations of effective temperatures and surface gravities. 

In a parallel work, \citet{El-Badry2018} systematically studied the impact of unresolved binary companions on the stellar parameters inferred by typical spectroscopic-survey pipelines assuming single stars. They found that, depending on the resolution, wavelength coverage, and signal-to-noise ratio of a particular survey, the unresolved binaries (which, we recall, may constitute $\sim50\%$ of the distant solar-type stars) produce systematic effects of the order of $300$ K in $T_{\rm eff}$, 0.1 dex in $\log g$, and 0.1 dex in [Fe/H]. Similar results were recently obtained by \citet{Lach2026} in a study dedicated to the effect of unresolved binaries in GALAH DR4 \citep{Buder2025}.

Combined, these two findings suggest that basically all of the current age catalogues used by the Galactic archaeology community are plagued by a significant amount of binary-induced systematic errors, on top of the age uncertainties coming from the uncertainties in stellar evolutionary models (see many other contributions to these proceedings). 
The scale of theses systematics depends of course on the characteristics of the sample (nearby vs. far, dwarfs vs. subgiants vs. giants), but the effects start to matter, especially when we interpret smaller subsamples. And this is {\it before} we even consider the biases induced by interacting-binary products that also affect a significant fraction of stars, both in clusters and in the field (e.g. \citealt{Chiappini2015, Fuhrmann2017a, Fuhrmann2026, Mathieu2025}).

\section{Beyond the single-star assumption: a new advance}

\begin{figure}
  \includegraphics[width=\textwidth]{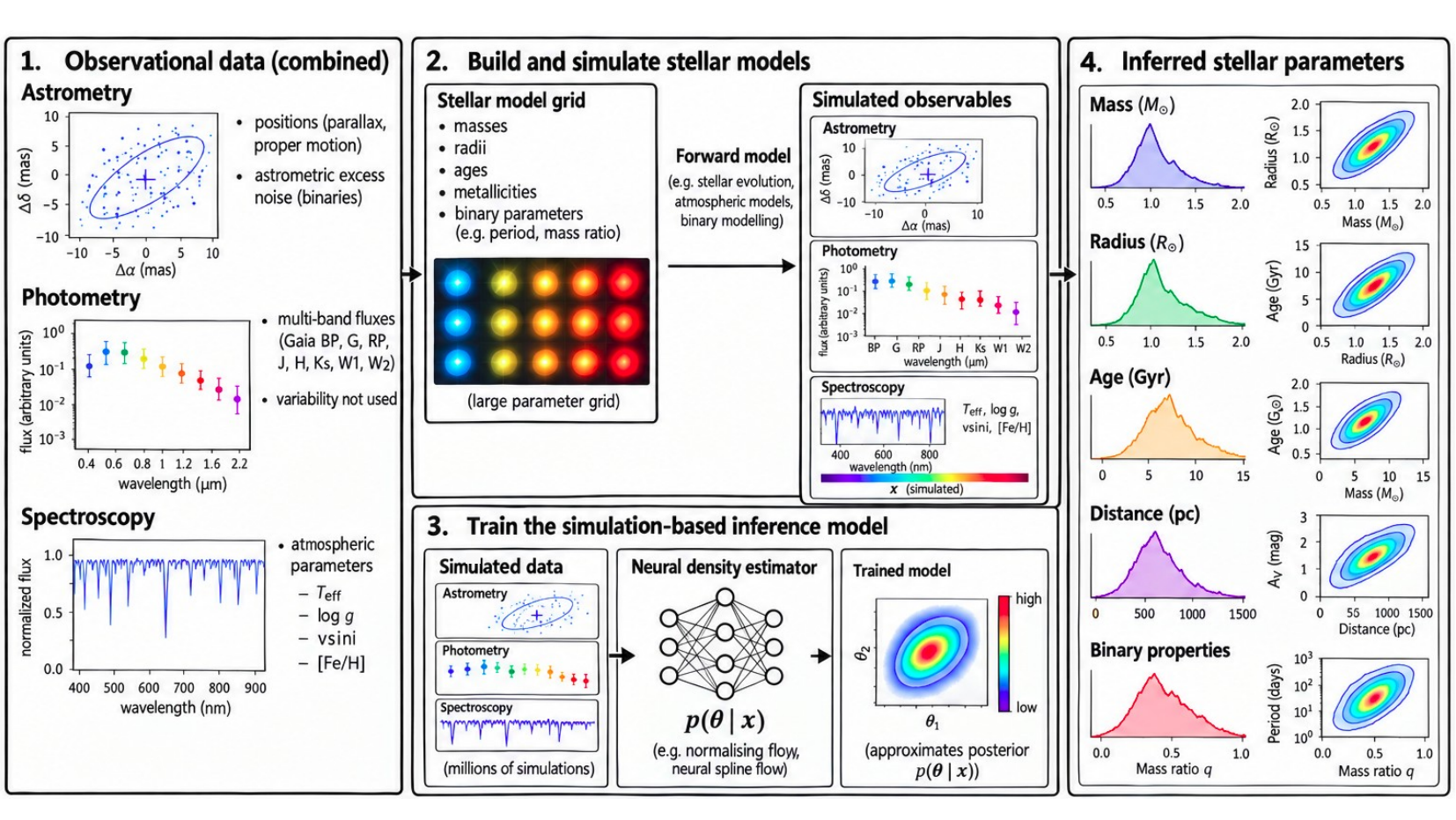}
  \caption{A schema of the simulation-based inference methodology applied to the problem of stellar-parameter inference from spectroscopic, photometric, and astrometric data, as pioneered in Malhotra et al. (in prep.).}
  \label{fig:sagar_paper_flowchart}
\end{figure}

In a recent paper, \citet{Malhotra2026} have started accounting for the effects of unresolved binaries in isochrone-fitting of open clusters. In particular, they developed a fast and robust framework, based on simulation-based inference (SBI; \citealt{Cranmer2020}), to jointly derive stellar masses and mass ratios of unresolved binaries, together with optimal open-cluster parameters. Once trained, the parameter inference with SBI was extremely fast (they 10,000 posterior samples per second on a single CPU). 

\begin{figure}
  \vspace{1cm}
  \includegraphics[width=\textwidth]{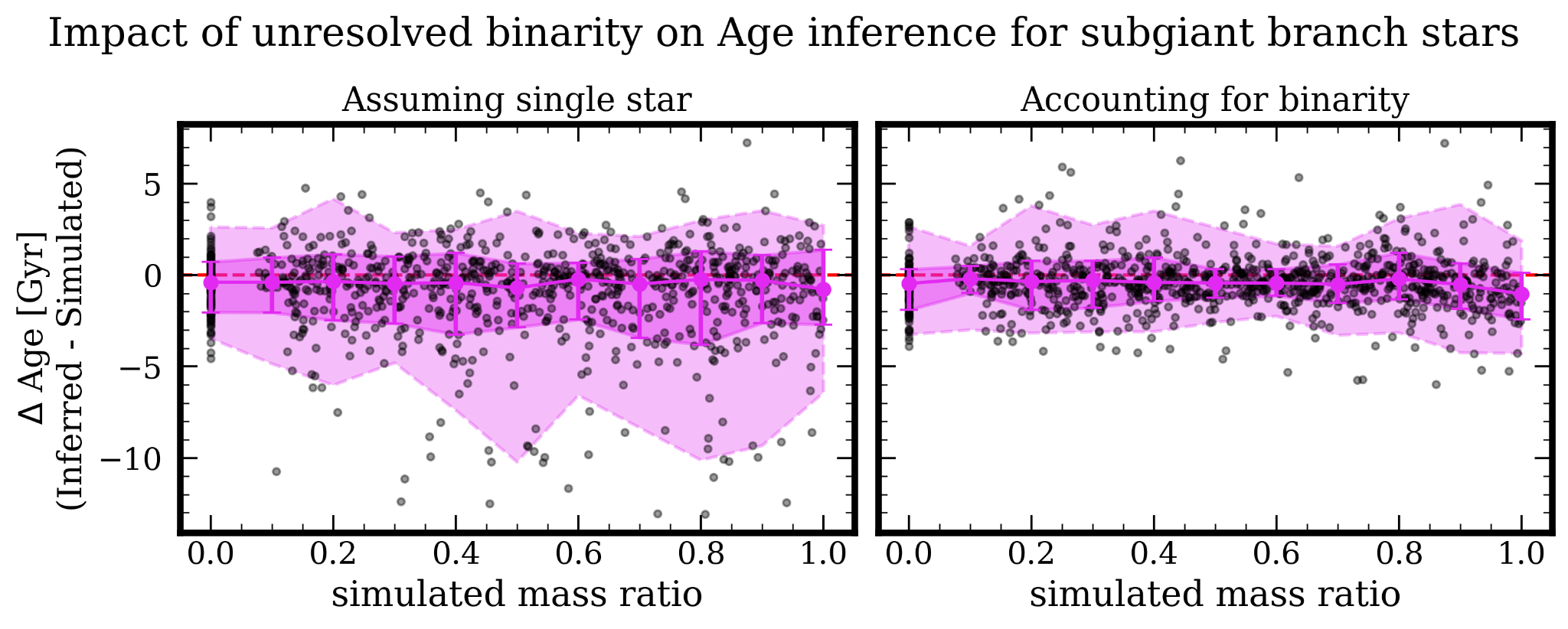}
  \caption{SBI age inference for simulated subgiant-branch stars using two models: assuming only single stars (left) and allowing for unresolved binarity (right). See Malhotra et al. (in prep.).}
  \label{fig:sagar_proofofconcept}
\end{figure}

Malhotra et al. (in prep.) are now following up on this initial experiment and extend their modelling to the case of field stars observed by spectroscopic surveys (e.g. GALAH), {\it Gaia} astrometry (including the astrometric error {\tt ruwe}), and multi-band photometric data. Figure \ref{fig:sagar_paper_flowchart} illustrates how SBI works in this case: we first simulate all plausible combinations of fundamental stellar parameters (mass, age, metallicity, rotation, binarity) on a massive grid and simulate what their astrometric, photometric, and spectroscopic 
parameters look like, applying realistic error models. Then we train a neural network to infer the input parameters from the simulated data. Once this training is done and sufficiently validated with unseen training data, we can infer the full posterior distribution for actually observed GALAH stars extremely efficiently (10,000 posterior samples per second on a single CPU). Among the preliminary findings is that accounting for unresolved binaries seems to not only improve the precision, but also the accuracy of the inferred ages of main-sequence turn-off stars (see Fig. \ref{fig:sagar_proofofconcept}).

\section{Summary}

We summarise our findings in the following statements:
\begin{enumerate}
    \item In stellar parameter inference, almost everything is correlated with age.
    \item Classical Bayesian inference is not dead - but becoming very expensive for large-scale surveys (especially without spectroscopic data).
    \item Stellar ages more precise than 10\% are within reach for hundreds of thousands of turnoff stars when combining {\it Gaia} and upcoming spectroscopic and photometric surveys.
    \item The most important age biases in large-scale surveys currently come from unresolved binarity (but also systematic uncertainties in stellar physics and interstellar extinction).
    \item We need higher-dimensional parameter inference to systematically address these biases.
    \item We need the next generation of stellar models calibrated on PLATO \citep{Rauer2025} and HAYDN \citep{Miglio2021} data to reach $1-5\%$ age accuracy.
\end{enumerate}

\subsection*{Acknowledgements}
\footnotesize{FA thanks the organisers and fellow participants of IAUS408 for the excellent symposium and the time in Liège. Following the suggestion of an unnamed participant, Fig. \ref{fig:sagar_paper_flowchart} in this proceeding was produced by a rather innocent LLM (GPT-5.6 Luna), although not-so-innocent readers might find this annoying.

FA acknowledges funding from MCIN/AEI/10.13039/501100011033 through grant RYC2021-031638-I, co-funded by European Union NextGenerationEU/PRTR. This work was partially supported by the Spanish MICIN/AEI/10.13039/501100011033 and by "ERDF A way of making Europe" by the European Union through grant PID2021-122842OB-C21 and PID2024-157964OB-C21, and the Institute of Cosmos Sciences University of Barcelona (ICCUB, Unidad de Excelencia Mar\'{\i}a de Maeztu) through grant CEX2024-001451-M and the project 2021-SGR-00679 GRC de l'Agència de Gestió d'Ajuts Universitaris i de Recerca (Generalitat de Catalunya).}

\bibliographystyle{aa}
\bibliography{Sample}

\end{document}